\documentclass{aastex}
\usepackage{emulateapj5}

\newcommand{\gtilde}
 {~ \raisebox{-1ex}{$\stackrel{\textstyle >}{\sim}$} ~}
\newcommand{\ltilde}
 {~ \raisebox{-1ex}{$\stackrel{\textstyle <}{\sim}$} ~}
\def\ltsima{$\; \buildrel < \over \sim \;$}
\def\ltsim{\lower.5ex\hbox{\ltsima}}
\def\gtsima{$\; \buildrel > \over \sim \;$}
\def\gtsim{\lower.5ex\hbox{\gtsima}}

\submitted{To Appear in Astrophysical Journal Letters
(Received January 5, 2001; Accepted February 14, 2001)}

\begin{document}
\begin{flushright}
NAOJ-Th-Ap/01-3
\end{flushright}

\title{Diffuse Extragalactic Background Light versus Deep Galaxy
Counts in the Subaru Deep Field: Missing Light in the Universe?}

\author{
Tomonori Totani$^1$, 
Yuzuru Yoshii$^{2, 3}$,
Fumihide Iwamuro$^4$, Toshinori Maihara$^5$, and Kentaro Motohara$^6$}

\altaffiltext{1}{Theory Division, 
National Astronomical Observatory, Mitaka, Tokyo 181-8588,
Japan (E-mail: totani@th.nao.ac.jp)}
\altaffiltext{2}{Institute of Astronomy, School of Science,
The University of Tokyo, 2-21-1 Osawa, Mitaka, Tokyo 181-8588, Japan}
\altaffiltext{3}{Research Center for the Early Universe, School of Science,
The University of Tokyo, Tokyo 113-0033, Japan}
\altaffiltext{4}{Department of Physics, Kyoto University, Kitashirakawa,
Kyoto 606-8502, Japan}
\altaffiltext{5}{Department of Astronomy, Kyoto University, Kitashirakawa,
Kyoto 606-8502, Japan}
\altaffiltext{6}{Subaru Telescope, National Astronomical Observatory of Japan,
650 North A'ohoku Place, Hilo, HI 96720, USA}

\shorttitle{DIFFUSE EBL VERSUS GALAXY COUNTS IN THE SUBARU DEEP FIELD}
\shortauthors{TOTANI ET AL.}

\date{\today}

\begin{abstract}
Deep optical and near-infrared galaxy counts are utilized to estimate
the extragalactic background light (EBL) coming from normal galactic light
in the universe. Although the slope of number-magnitude relation
of the faintest counts is flat enough for the count integration to
converge, considerable fraction of EBL from galaxies could still have been
missed in deep galaxy surveys because of various selection effects
including the cosmological dimming of surface brightness of galaxies.
Here we give an estimate of EBL from galaxy counts, in which these
selection effects are quantitatively taken into account for the
first time, based on reasonable models of galaxy evolution
which are consistent with all available data of galaxy counts, size, and
redshift distributions. We show that the EBL from galaxies is best
resolved into discrete galaxies in the near-infrared bands ($J, K$)
by using the latest data of the Subaru Deep Field; more than 80-90\%
of EBL from galaxies has been resolved in these bands.
Our result indicates that the contribution by missing galaxies cannot account
for the discrepancy between the count integration and recent tentative
detections of diffuse EBL in the $K$-band (2.2 $\mu$m), 
and there may be a very diffuse
component of EBL which has left no imprints in known galaxy populations.
\end{abstract}


\section{Introduction}
Extragalactic background light (EBL) in the optical and near-infrared (NIR)
wavebands is a fundamental quantity for galaxy formation and cosmology,
which is believed to be dominated by the integration of all stellar
light in the universe (Bond, Carr, \& Hogan 1986; Yoshii \& Takahara 1988). 
If all stellar light is emitted from galactic systems, 
the EBL can be resolved into discrete galaxies by deep galaxy surveys.
The deepest image of the universe in the optical bands
has been obtained by the Hubble Deep Field (HDF; Williams et al. 1996).
The faint-end slopes of the HDF galaxy counts in all the four optical
bands ($U_{300}$, $B_{450}$, $V_{606}$, and $I_{814}$) are flatter than
the critical slope index $d(\log N)/(dm) = 0.4$, with which the
contribution of galaxies to the EBL is constant against magnitudes.
Therefore the extrapolation of the galaxy counts into fainter
magnitudes does not significantly increase EBL but converges
to a finite EBL flux, and this means that
the bulk of EBL from galactic light has already been resolved into
discrete galaxies (e.g., Madau \& Pozzetti 2000). 
The situation is the same for the NIR 
band, although there has been a considerable scatter in 
the faint-end counts in the $K$ band.  In fact, the latest
$K$ count data of the Subaru Deep Field (SDF; Maihara et al. 2000) with 
350 galaxies down to the 5$\sigma$ limiting magnitude of $K'=23.5$
show a very flat slope of $d(\log N)/dm \sim$ 0.23 to $K \sim $ 24.
 
These results of faint galaxy counts therefore require that the diffuse
EBL in optical and NIR bands should not be different from the count
integrations, provided that the ordinary galactic light is the dominant
source of the EBL in these bands, as generally believed. 
However, recent (tentative) detections
of diffuse EBL in these bands suggest that the diffuse EBL flux is
consistently higher than the count integrations.
The measurement of the optical EBL by Bernstein et al. (1999)
is higher than the optical count integrations by Madau \& Pozzetti
(2000) by a factor of $\sim 2$--4. There are several independent
reports for detection of the diffuse EBL
at the $K$ band (2.2 $\mu$m): $\nu I_\nu = 22.4 \pm 6 
\ \rm nW \ m^{-2} sr^{-1}$ (Gorjian, Wright, \& Charly 2000),
$20.2 \pm 6.3$ (Wright 2001), and $29.3 \pm 5.4$ (Matsumoto et al. 2000),
which should be compared with the integration of
$K$ counts ($\sim 8  \ \rm nW \ m^{-2} sr^{-1}$, Madau \& Pozzetti 2000).

It should be noted that this is 
a comparison between two purely observable quantities, and no theoretical
modeling is included. Any theoretical model of galaxy formation 
cannot reproduce {\it simultaneously} the counts and EBL, although
it is rather easy to construct a model to explain either of the two.
If the discrepancy between the diffuse EBL and count integration is
real, it might suggest the existence of very diffuse component which
is different from normal galaxies.
Before deriving this extraordinary conclusion, however, all possible
systematic uncertainties in the above estimates must extensively be
checked. One of such systematics is the contribution to EBL by the galaxies
missed in deep galaxy surveys. Since galaxies are extended sources,
the detectability near the detection limit is not as simple as 
point sources. Furthermore, the well-known effect of the cosmological
dimming of surface brightness [$S \propto (1+z)^{-4}$] should make 
high-$z$ galaxies very difficult to detect, while such objects
may have a significant contribution to EBL. The photometry scheme could
also be a problem, because there is considerable uncertainty
in the estimate of the magnitude of faint galaxies because of
`growing' the photometry beyond the outer detection isophotes of galaxies.

In spite of this importance,
no realistic or quantitative estimate of the contribution to EBL from
these missing galaxies has been made so far.
The purpose of this Letter is to make such an estimate, using realistic
galaxy evolution models which reproduce the local galaxy populations
as well as deep counts, size and redshift 
distributions of the faintest galaxies.
We will calculate how much galactic light is missed in current deep surveys
taking into account the effects mentioned above, under the
observational conditions and detection criteria of HDF for optical
bands and those of SDF for NIR bands. Then we will derive our best-guess
for the EBL flux coming from normal galaxies.

\section{Method and Model}
\label{section:method}
The examined systematic effects which could lead to missing of faint galaxies
are as follows: (1) apparent size and
surface brightness profiles of galaxies where the cosmological
dimming is taken into account, (2) dimming of an image by seeing,
(3) criteria and completeness of galaxy detections under the
observational conditions, and
(4) photometric scheme (isophotal magnitude applied consistently).

We estimate the contribution of missing galaxies to EBL as follows.
First, we construct a model of galaxy counts which best fits to the observed
counts, taking into account all the above selection effects.
Then we can calculate the true galaxy counts and 
EBL flux using the same model without selection effects, and comparison
between the true counts and observed counts gives an estimate of
contribution by missing galaxies. The general formalism to include
the selection effects in calculation of galaxy counts has been given
in Yoshii (1993), and we have already analyzed the HDF counts and 
photometric redshift distributions by this method (Totani \& Yoshii
2000, hereafter TY00). 
Here we briefly summarize the methods and the model of TY00.

The number density
of galaxies is normalized at $z$=0 by the observed $B$ band luminosity
function (see Table 1 of TY00).
Galaxies are classified into five morphological types of
E/S0, Sab, Sbc, Scd, and Sdm, and their luminosity evolution is
followed by luminosity evolution models of Arimoto \& Yoshii 
(1987) and Arimoto, Yoshii, \& Takahara (1992). These models are made
to reproduce colors and chemical properties
of local galaxies. Absorptions by interstellar dust and intergalactic
HI clouds are taken into account. 
Possible number evolution of galaxies is considered by a phenomenological model
in which the Schechter parameters of the luminosity function
have a redshift dependence as $\phi^* \propto (1+z)^\eta$
and $L^* \propto (1+z)^{-\eta}$, i.e., luminosity density is 
conserved. The formation redshift $z_F$ of galaxies is simply assumed
to be 5 for all galaxies, 
but changing this parameter in a reasonable range ($z_F \sim$ 3--10)
would hardly change the conclusion
derived in this letter. (See Table 3 of TY00 for the summary of 
dependence of the count model on 
various parameters and systematic model uncertainties.)

The selection effects are calculated using
simplified one-dimensional smooth-profile models 
and analytic estimates of the surface brightness thresholds.
This is a good first approximation, although 
our calculation does not include a full simulation
of the images and their noise properties with
two-dimensional galaxy morphologies. The size of galaxies is
estimated by the
size-luminosity relation observed for local galaxies. The exponential
and de Vaucouleurs' laws are assumed for the surface brightness profile 
of spiral and elliptical galaxies, respectively. 
The galaxy size is assumed to be constant
except in the case of number evolution. We assume that the change of galaxy 
luminosity and size caused by number evolution
obeys a scaling relation of $L \propto r^\xi$
during the merger processes with $\xi=3$, and changing the value of
$\xi$ in a reasonable range of 2--4 hardly affects the count predictions,
unless unrealistically strong number evolution is invoked (TY00).
The possible intrinsic size evolution not induced by merging will be
checked in \S \ref{section:check}. We calculate the isophotal size
and isophotal magnitude
for each model galaxy using the detection isophote applied
in the HDF and SDF surveys,
taking into account all the selection effects mentioned above.
The galaxies meeting the detection criteria of the HDF or SDF are
counted in the model galaxy counts,
and then they are compared to the raw observed counts consistently
as a function of isophotal magnitudes. In this way we find 
the best model to explain the observations with the selection effects
properly taken into account.

\section{Results}
\begin{figure*}
\centerline{\epsfxsize=9cm \epsfbox{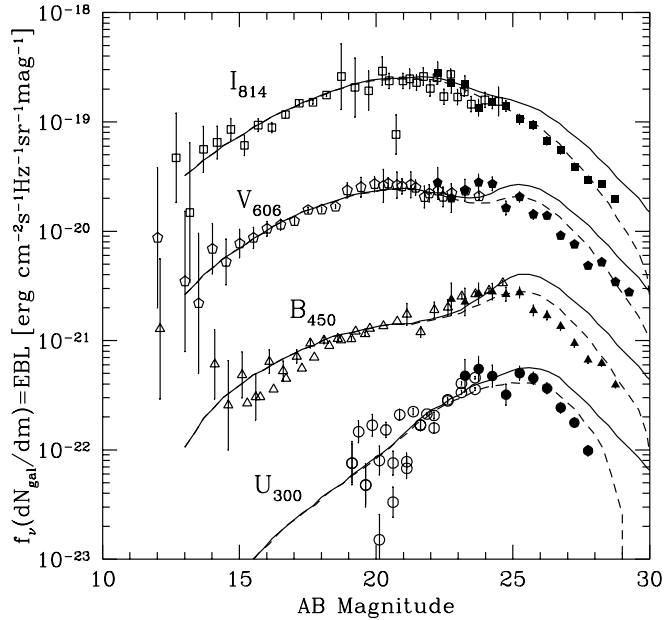}}
\caption{
The contribution to EBL by galaxies in the four
optical bands as functions
of AB magnitudes. 
The filled symbols are the HDF data (in isophotal magnitudes), while the
open symbols are ground-based data (see TY00 for references).
The dashed lines are the predictions by the model A (see text)
taking into account the
selection effects under the observational conditions of HDF,
fitting to the observed counts. The solid lines are the prediction by
the same model, but without the selection effects in total magnitudes.
For clarity, the data and model curves of the $B_{450}$, $V_{606}$,
and $I_{814}$ bands are multiplied by factors of $10^{0.5}$, $10^{1.3}$,
and $10^{2}$, respectively.
}
\label{fig:EBL-HDF}
\end{figure*}

TY00 found that
a galaxy count model with a modest number evolution ($\eta = 1$)
in a $\Lambda$-dominated flat universe with $(\Omega_0, \Omega_\Lambda,
h) = (0.2, 0.8, 0.7)$ gives the best explanation not only for the observed
counts but also the photometric redshift distribution of the HDF galaxies.
This cosmological model is now the best favored by various cosmological
observations.
The number evolution of $\eta \sim 1$ is also consistent with the
observational constraint on the merging fraction of galaxies
at $z \ltilde 1$ (Le F\'evre et al. 2000). This model is referred to
as the model A in the following.

Figure \ref{fig:EBL-HDF} is the galaxy counts multiplied by flux,
showing the contribution to EBL in the four 
passbands of the HDF as functions of AB
magnitudes. The dashed lines are the predictions of the model A with the
selection effects taken into account, and they exhibit a reasonable
agreement with the observed counts in isophotal magnitudes. 
On the other hand, the solid line
shows the model predictions in total magnitudes
without the selection effects, i.e., the
true galaxy counts. The excess of the solid lines over the dashed lines
gives an estimate of the contribution by the missing galaxies to the
EBL. 

\begin{figure*}
\centerline{\epsfxsize=9cm \epsfbox{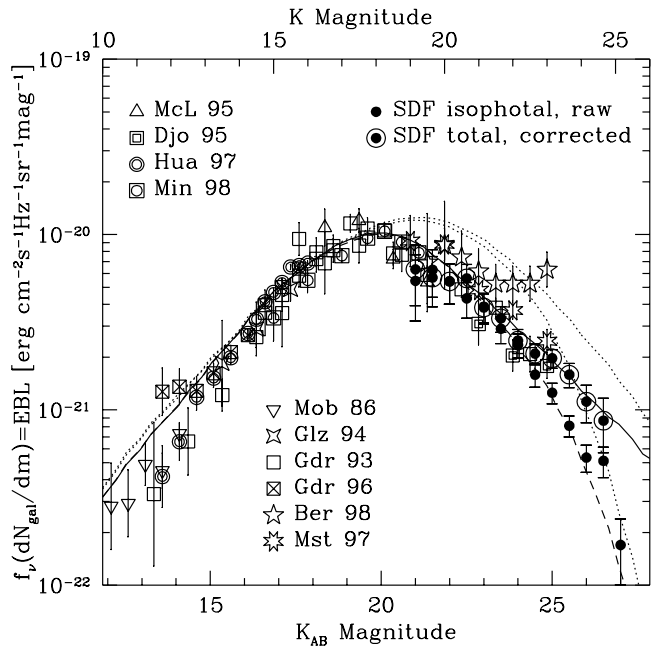}}
\caption{The same as Fig. \ref{fig:EBL-HDF}, but for the
$K$ band. 
The filled circles are the raw SDF counts in isophotal 
magnitude, while the symbols $\odot$ are the counts in total magnitude
which are corrected for incompleteness assuming point sources 
(Maihara et al. 2000). The dashed
line is the prediction by the model B (see text) 
for which the selection effects
under the observational conditions of SDF are taken into account,
fitting to the raw counts. The solid line
is the same prediction, but the selection effects are not included.
The two dotted lines are the prediction by model A which is used in
Fig. \ref{fig:EBL-HDF}, with and without the selection effects.
The other data are, McLeod et al. 1995 (McL95), Djorgovski et al.
1995 (Djo95), Huang et al. 1997 (Hua97), Minezaki et al. 1998 (Min98), 
Mobasher et al. 1986 (Mob86),
Glazebrook et al. 1994 (Glz94), Gardner et al. 1993, 1996 (Gdr93, Gdr96), 
Bershady et al. 1998 (Ber98),
and Moustakas et al. 1997 (Mst97).
}
\label{fig:EBL-SDF}
\end{figure*}

Figure \ref{fig:EBL-SDF} shows the contribution to EBL in the $K$ band,
including the latest data of the SDF. In Fig. \ref{fig:EBL-HDF}
we have used the model A with
number evolution of $\eta = 1$, but we found that this model seriously
overpredicts the $K$ counts, as shown by dotted lines in this figure.
Rather, the $K$ band counts
can be fitted better by the pure luminosity evolution model 
with no number evolution ($\eta = 0$), when
the same cosmological model as in the HDF is used (see Totani et al.
2001 in detail). The dashed line is this model with the selection
effects, fitting well to the observed isophotal raw counts of the SDF
(filled circles). The solid line is the prediction as a function of
total magnitudes without any selection effects. For comparison,
the SDF counts corrected for incompleteness assuming that all objects
are point sources (Maihara et al. 2000) are also shown by the
symbol $\odot$ as a function of total magnitudes.
Here we assumed $K = K' - 0.1$.

This discrepancy between optical and NIR counts 
is probably coming from the limitation
of the model assuming the same number evolution for all galaxy types.
In the $K$ band,
elliptical or early type galaxies are more dominant in number compared with
the optical bands. Therefore, this result may be understood if there is
no or weaker number evolution for elliptical galaxies than
that for other types. In addition, 
the giant and dwarf elliptical galaxies have been treated as
distinct populations in Fig. \ref{fig:EBL-SDF}, 
because it fits to the faintest $K$ counts
even better (see Totani et al. 2001 for the details). We refer to this model
as the model B hereafter. We will use both of the two models in estimating
the EBL, to see the model dependence of our calculation.

Now the contribution to EBL from galaxies missed in HDF and SDF can
be estimated.
We estimate the true galaxy counts by using the observed 
galaxy counts and models as follows:
\begin{eqnarray}
N_{\rm true}(m) = \left\{
\begin{array}{ll}
N_{\rm obs}(m) \left( \frac{N_{\rm m1}(m)}{N_{\rm m2}(m)} \right) , &
(m < m_{\lim}) \\
N_{\rm obs}(m_{\lim})
\left( \frac{N_{\rm m1}(m)}{ N_{\rm m2}(m_{\lim}) } \right) , &
(m > m_{\lim}) \\
\end{array}\right.
\end{eqnarray}
where $m_{\lim}$ is the faint limit of observed magnitude, 
$N_{\rm obs}$ is the observed counts, and $N_{\rm m1}$ and
$N_{\rm m2}$ are the model counts without/with the selection
effects, respectively. The estimate of EBL is simply given by
the integration of $N_{\rm true}$.
The ratio of the raw count integration to
the true EBL from galactic light, which we call a resolution fraction,
is shown in Table \ref{table:corrections} both for the model A and B.
The dependence on the two models is not significant. An overall trend is
that the resolution fraction becomes greater with increasing
wavelength, because the evolutionary effect of galaxies becomes less
significant. Therefore, the best evidence that the bulk of EBL from
galactic light has been
resolved into discrete galaxies is given by the $J$ and $K$ counts of
SDF; more than 80--90\% of the NIR galactic light in the universe
has been resolved.

\section{Checking Reliability of Our Results}
\label{section:check}
It should be noted that our estimate of the EBL flux from galaxies
is essentially based on the observed counts, and the uncertainty concerning the
model used here is relevant only to the contribution from missed galaxies.
Given that this contribution of our best guess is not large compared with that
from resolved counts, it is unlikely that the model uncertainty
drastically changes the estimate of total EBL flux from normal galaxies.

\begin{figure*}
\centerline{\epsfxsize=9cm \epsfbox{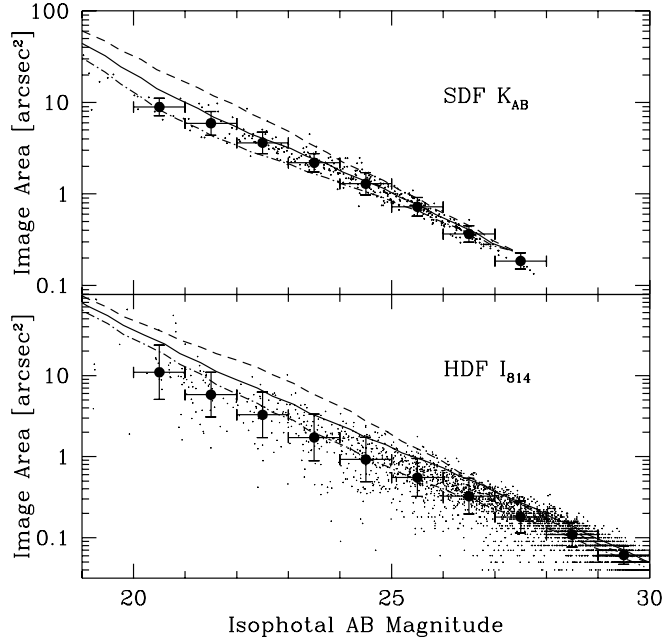}}
\caption{
Magnitude-size relation for SDF and HDF galaxies.
The size is estimated by the isophotal area of galaxies.
The small dots are for individual galaxies, while the filled circles
are mean size, with the vertical and horizontal error bars 
showing $1\sigma$ dispersion and the magnitude intervals, respectively.
Three curves are the mean isophotal area predicted by the 
theoretical model (model A for HDF and model B for SDF, see text),
for different size-evolutions of galaxies:
$r_e \propto (1+z)^\zeta$ with $\zeta =$ 1 (dashed), 0 (solid), 
and $-1$ (dot-dashed).
}
\label{fig:size}
\end{figure*}

To demonstrate the reliability of our analysis, 
we show a comparison of the observed isophotal size of galaxies
and that predicted by the model used here, 
in Fig. \ref{fig:size}, and this is a crucial check whether
we have successfully modeled the systematic selection effects.
In the above models we have assumed that the galaxy sizes do not evolve
intrinsically with time except in the case of mergers. 
Fig. \ref{fig:size} shows that this no-size-evolution model is in 
reasonable agreement with the data, especially in the SDF.
In order to check possible size evolution, 
we have also calculated the model prediction
with a simple intrinsic size evolution (i.e., not caused by mergers), 
as $r_e \propto
(1+z)^\zeta$ with $\zeta = $ $-1$ and 1, where $r_e$ is the effective 
radius of galaxies.
The $\zeta = -1$ model is favored rather than the no-size-evolution
model by the HDF galaxies, and hence we also show the resolution fraction of
this case in Table \ref{table:corrections}. The resolution fraction
becomes larger by the size evolution with $\zeta = -1$, 
because galaxies with smaller size are more easily detected
when luminosity is fixed. 

The discrepancy between HDF and SDF
size distributions is probably coming from dependence of size
evolution on galaxy types. There is no evidence for number or size
evolution for elliptical galaxies, while later type galaxies seem
to have evolved in size and number to some extent.
In either case, there is no evidence 
that the size of high-$z$ galaxies is intrinsically larger than local
galaxies ($\zeta > 0$), and hence it is very 
unlikely that the size evolution effect drastically increases
the contribution of missing galaxies to resolve the discrepancy
between counts and EBL.

We have used the empirical mean luminosity-size
relation of local galaxies in the calculation of galaxy sizes. 
However, there is considerable scatter along the
mean relation, and 
the largest uncertainty in the estimate of the selection effects
is probably coming from this scatter.
We have calculated the resolution fraction using the
size-luminosity relation shifted by +1$\sigma$ scatter in $\Delta (\log r_e)$
(see Fig. 3 of TY00), in the direction of larger
size and hence more significant selection effects. 
The results are given in Table \ref{table:corrections}, and
the resolution fraction in the $B$ band could be as small as
$\sim$ 60\% by this uncertainty, 
but that in the $K$ band is still $\gtilde$90\% (Table 
\ref{table:corrections}).
The resolution fraction of SDF is less sensitive than HDF to the uncertainties
about galaxy sizes, as can be seen in Table \ref{table:corrections}.
This is because the seeing size is comparable or larger than the
original galaxy sizes. In fact, 
the solid line in Fig. \ref{fig:EBL-SDF} is very close
to the count data independently corrected assuming point sources ($\odot$), and
this suggests that the extended nature of galaxies is not important
at least in SDF. [Note that the situation is different for HDF
in which the cosmological dimming is important (TY00), 
because of the better angular resolution than SDF.]
This result further reduces the room of model uncertainties
in our estimate of missing galactic light in the $K$ band.

\section{Conclusions}
\label{section:discussion}
For the first time we presented a quantitative estimate of contribution
to EBL from galaxies missed in deep surveys by various selection effects.
The range of true EBL from galaxies of our best-guess is shown in Table 
\ref{table:corrections}, considering the range of resolution fractions
of the four models and uncertainties in the raw integration of observed
counts. Then the $K$ band (2.2$\mu$m) EBL flux from all galaxies 
is unlikely to be larger than 
$10.2 \ \rm nW \ m^{-2} sr^{-1}$, but this is considerably
smaller than the recent direct measurements of EBL in this band
shown in Table \ref{table:corrections}.

Therefore, normal galaxies missed in deep surveys cannot reconcile the
discrepancy between the count integration and diffuse EBL in optical
and NIR bands.
Unless there is crucial systematic error in the diffuse EBL measurements,
there must be a very diffuse component of EBL
which cannot be explained by known galaxy populations. If it is the case, 
the impact on galaxy formation and cosmology would
be quite significant, and further study is required especially
for more accurate measurements of diffuse EBL.

This work is partially
based on the data corrected at the Subaru telescope, which is operated
by the National Astronomical Observatory of Japan.
We would like to thank T. Matsumoto for providing us with his data of diffuse
EBL measurement.

{\it Note Added.---} After acceptance of this letter, we found
two more reports for the $J$ and $K$ band EBL flux, by
Write \& Reese (2000) and Cambresy et al. (2001). These
are added in the reference lists as well as in Table 1.
Both are well consistent with the other diffuse EBL measurements,
and it seems that the discrepancy
between counts and diffuse EBL is even more severe in the J band
than the K band.

\begin{table}
\footnotesize
\begin{center}
\caption{EBL Flux and Resolution Fraction of Resolved Galaxies}
\label{table:corrections}
\begin{tabular}{lcccccc}
\hline \hline
Band ($\lambda/$\AA) 
& U(3000) & B(4500) & V(6100) & I(8100) & J(12500) & K(22000) \\
\hline
& \multicolumn{6}{c}{ Simple Integration of Observed Galaxy Counts$^a$ }\\
Ref. 1              & $(2.9^{+0.6}_{-0.4})^b$ & $4.6^{+0.7}_{-0.5}$ 
                    & $(6.7^{+1.3}_{-0.9})^b$ & $8.0^{+1.6}_{-0.9}$
                    & $(9.7^{+3.0}_{-1.9})^b$ & $7.9^{+2.0}_{-1.2}$ \\
This work & $2.7 \pm 0.3$ & $4.4 \pm 0.4$  & $6.0 \pm 0.6$ 
          & $8.1 \pm 0.8$ & $10.9 \pm 1.1$ & $8.3 \pm 0.8$ \\
\hline
& \multicolumn{6}{c}{ Estimate of Resolution Fractions }\\
Model A$^c$                       & 0.78 & 0.78 & 0.87 & 0.90 & 0.97 & 0.93 \\
Model B$^c$                       & 0.82 & 0.83 & 0.90 & 0.92 & 0.95 & 0.92 \\
size evolution ($\zeta=-1$)$^d$   & 0.81 & 0.92 & 0.93 & 0.96 & 0.94 & 0.96 \\
$+1\sigma$ in size$^d$            & 0.68 & 0.61 & 0.76 & 0.82 & 0.95 & 0.89 \\
\hline
& \multicolumn{6}{c}{ Our Best-Guess EBL from All Normal Galaxies in the
Universe$^a$}\\
& 2.9--4.4 & 4.3--7.9 & 5.8--8.9 & 7.6--10.9 & 10.1--12.8 & 7.8--10.2 \\
\hline
 & \multicolumn{6}{c}{ Measurements of Diffuse EBL$^a$ } \\
Ref. 2 & 12.0$\pm$5.7 &  & (14.9$\pm$4.4)$^b$ & 17.6$\pm$4.8 & & \\
Ref. 3 & & & & &                 & 22.4$\pm$6.0 \\
Ref. 4 & & & & & 28.9$\pm$16.3   & 20.2$\pm$6.3 \\
Ref. 5 & & & & & 60$\pm$15       & 29.3$\pm$5.4 \\
Ref. 6 & & & & &                 & 23.1$\pm$5.9 \\
Ref. 7 & & & & & 54.0$\pm$16.8   & 27.8$\pm$6.7 \\
\hline
\hline
\end{tabular}
\tablenotetext{}{$^a$ \ The EBL and count integration 
$\nu I_\nu$ in units of $\rm nW \ m^{-2} \ sr^{-1}$.}

\tablenotetext{}{$^b$ Estimated at slightly different wavelengths
from ours (see original references).}

\tablenotetext{}{$^c$ Model A: number evolution with $\eta =1$. 
Model B: no number evolution with $\eta =0$, and distinct treatment for
giant and dwarf elliptical galaxies. (See text for detail.)
}

\tablenotetext{}{$^d$ Using models A and B for 3000-8100 and
12500--22000 {\AA}  bands, respectively.}

\tablenotetext{}{References. --- (1) Madau \& Pozzetti (2000),
(2) Bernstein et al. (2001),
(3) Gorjian, Wright, \& Chary (2000), (4) Wright (2001), 
(5) Matsumoto et al. (2000), (6) Wright \& Reese (2000), and
(7) Cambresy et al. (2001)
}
\end{center}
\end{table}

\end{document}